\documentclass[a4paper,11pt]{article}
\pdfoutput=1 

\usepackage{jcappub} 

\usepackage[T1]{fontenc} 

\usepackage{graphicx}  
\usepackage{siunitx}  
\usepackage{epsfig} 
\usepackage{subfigure}
\usepackage{lineno}  
\usepackage{amsmath,amssymb}
\usepackage{comment}
\usepackage{multirow}
\usepackage{dcolumn}
\usepackage{bm}
\usepackage{setspace}
\usepackage{multibib}
\includecomment{printrobustnotes}
\includecomment{printallnotes}
\usepackage{xspace}
\usepackage{color}
\def\qna{QF$_{\mathrm{Na}}$~}
\def\qnadot{QF$_{\mathrm{Na}}$}
\def\qi{QF$_{\mathrm{I}}$~}
\def\qidot{QF$_{\mathrm{I}}$}

\title{\boldmath Comparison between DAMA/LIBRA and COSINE-100 in the light of Quenching Factors}

\author{The COSINE-100 Collaboration\hspace{15em}}
\author[a]{Y.~J.~Ko}
\author[b]{G.~Adhikari}
\author[b,1]{P.~Adhikari}
\author[c]{E.~Barbosa~de~Souza}
\author[d]{N.~Carlin}
\author[e]{J.~J.~Choi}
\author[e]{S.~Choi}
\author[f]{M.~Djamal}
\author[g]{A.~C.~Ezeribe}
\author[a]{C.~Ha}
\author[h]{I.~S.~Hahn}
\author[a]{E.~J.~Jeon}
\author[c]{J.~H.~Jo}
\author[a]{W.~G.~Kang}
\author[j]{M.~Kauer}
\author[k]{G.~S.~Kim}
\author[a]{H.~Kim}
\author[k]{H.~J.~Kim}
\author[a]{K.~W.~Kim}
\author[a]{N.~Y.~Kim}
\author[e]{S.~K.~Kim}
\author[a,b,m]{Y.~D.~Kim}
\author[a,l,m]{Y.~H.~Kim}
\author[a]{E.~K.~Lee}
\author[a,m]{H.~S.~Lee}
\author[a]{J.~Lee}
\author[k]{J.~Y.~Lee}
\author[a,m]{M.~H.~Lee}
\author[m,a]{S.~H.~Lee}
\author[a]{D.~S.~Leonard}
\author[g]{W.~A.~Lynch}
\author[d]{B.~B.~Manzato}
\author[c]{R.~H.~Maruyama}
\author[g]{R.~J.~Neal}
\author[a]{S.~L.~Olsen}
\author[m,a]{B.~J.~Park}
\author[n]{H.~K.~Park}
\author[l]{H.~S.~Park}
\author[a]{K.~S.~Park}
\author[d]{R.~L.~C.~Pitta}
\author[f]{H.~Prihtiadi}
\author[a]{S.~J.~Ra}
\author[i]{C.~Rott}
\author[a]{K.~A.~Shin}
\author[g]{A.~Scarff}
\author[g]{N.~J.~C.~Spooner}
\author[c]{W.~G.~Thompson}
\author[o]{L.~Yang}
\author[i]{and G.~H.~Yu\note{Present address : Department of Physics, Carleton University, Ottawa, ON K1S 5B6, Canada}}
\author{\hspace{15em}}
\author{\hspace{15em}}

\affiliation[a]{Center for Underground Physics, Institute for Basic Science (IBS), Daejeon 34126, Republic of Korea}
\affiliation[b]{Department of Physics, Sejong University, Seoul 05006, Republic of Korea}
\affiliation[c]{Department of Physics and Wright Laboratory, Yale University, New Haven, CT 06520, USA}
\affiliation[d]{Physics Institute, University of S\~{a}o Paulo, 05508-090, S\~{a}o Paulo, Brazil}
\affiliation[e]{Department of Physics and Astronomy, Seoul National University, Seoul 08826, Republic of Korea}
\affiliation[f]{Department of Physics, Bandung Institute of Technology, Bandung 40132, Indonesia}
\affiliation[g]{Department of Physics and Astronomy, University of Sheffield, Sheffield S3 7RH, United Kingdom}
\affiliation[h]{Department of Science Education, Ewha Womans University, Seoul 03760, Republic of Korea} 
\affiliation[i]{Department of Physics, Sungkyunkwan University, Suwon 16419, Republic of Korea}
\affiliation[j]{Department of Physics and Wisconsin IceCube Particle Astrophysics Center, University of Wisconsin-Madison, Madison, WI 53706, USA}
\affiliation[k]{Department of Physics, Kyungpook National University, Daegu 41566, Republic of Korea}
\affiliation[l]{Korea Research Institute of Standards and Science, Daejeon 34113, Republic of Korea}
\affiliation[m]{IBS School, University of Science and Technology (UST), Daejeon 34113, Republic of Korea}
\affiliation[n]{Department of Accelerator Science, Korea University, Sejong 30019, Republic of Korea}
\affiliation[o]{Department of Physics, University of Illinois at Urbana-Champaign, Urbana, IL 61801, USA}

\emailAdd{yjko@ibs.re.kr}
\emailAdd{kwkim@ibs.re.kr}
\emailAdd{hyunsulee@ibs.re.kr}

\abstract{ There is a long standing debate about whether or not the annual modulation signal reported by the DAMA/LIBRA
  collaboration is induced by Weakly Interacting Massive Particles~(WIMP) in the galaxy's dark matter halo
  scattering from nuclides in their NaI(Tl) crystal target/detector. This is because regions of
  WIMP-mass vs. WIMP-nucleon cross-section parameter space that can accommodate the DAMA/LIBRA-phase1
  modulation signal in the context of the standard WIMP dark matter galactic halo and
  isospin-conserving~(canonical), spin-independent~(SI) WIMP-nucleon interactions have been excluded by many of other dark matter search experiments 
	including COSINE-100, which uses the same NaI(Tl) target/detector material.
  Moreover, the recently released DAMA/LIBRA-phase2 results are inconsistent with an interpretation as WIMP-nuclide
  scattering via the canonical SI interaction and prefer, instead, isospin-violating or spin-dependent interactions.
  Dark matter interpretations of the DAMA/LIBRA signal are sensitive to the NaI(Tl) scintillation efficiency for
  nuclear recoils, which is characterized by so-called quenching factors~(QF), and the QF values used in previous
  studies differ significantly from recently reported measurements, which may have led to incorrect interpretations
  of the DAMA/LIBRA signal. In this article, the compatibility of the DAMA/LIBRA and COSINE-100 results,
  in light of the new QF measurements is examined for different possible types of WIMP-nucleon interactions. The
  resulting allowed parameter space regions associated with the DAMA/LIBRA signal are explicitly
  compared with 90\% confidence level upper limits from the initial 59.5~day COSINE-100 exposure. With the newly
  measured QF values, the allowed 3$\sigma$ regions from the DAMA/LIBRA data are still generally excluded by
  the COSINE-100 data.  
}

\begin{document}
\maketitle
\flushbottom
\section{Introduction}
A number of astrophysical observations provide evidence that the dominant matter
component of the universe is not ordinary matter, but rather non-baryonic dark
matter~\cite{Clowe:2006eq,Aghanim:2018eyx}. Weakly Interacting Massive
Particles~(WIMPs) are particle dark matter
candidates~\cite{PhysRevLett.39.165,Jungman:1995df,Goodman:1984dc} that have been
the subject of extensive searches by direct detection, indirect detection,
and collider experiments, with no success~\cite{Tanabashi:2018oca}. 

The one exception is the long-standing observation by the DAMA/LIBRA collaboration of an 
annual modulation in the low-energy event rate in an underground array of low-background NaI(Tl)
detectors. Although this signal has persisted throughout more than 20 years of
investigation~\cite{Bernabei:1998fta,Bernabei:2008yi,Bernabei:2010mq,Bernabei:2013xsa,Bernabei:2018yyw},
its interpretation as being due to WIMP-nucleus scattering in the specific context of the standard
galactic WIMP halo model~\cite{Lewin:1995rx,freese1987}, has been the subject of a continuing debate.
This is because the WIMP-nucleon cross sections inferred from the DAMA/LIBRA modulation are in conflict
with limits from other experiments that directly measure the total, time integrated rate of nuclear
recoils~\cite{Kim:2012rza,Angloher:2014myn,Agnese:2014aze,Abe:2015eos,PhysRevLett.118.021303,Aprile:2017yea,Agnese:2017njq,Aprile:2018dbl,Agnes:2018ves,Akerib:2018zoq,Abdelhameed:2019hmk}.
An unambiguous verification of the DAMA/LIBRA signal by independent experiments using the same NaI(Tl) crystal
target material is mandatory. Experimental efforts by several groups using the same NaI(Tl) target medium are
currently underway~\cite{Kim:2014toa,sabre,Adhikari:2017esn,Fushimi:2018qzk,Coarasa:2018qzs,Amare:2018sxx}. 

COSINE-100, located at the Yangyang underground laboratory in South Korea, is one of the experiments aimed
at testing the DAMA/LIBRA results with a NaI(Tl) crystal detector/target~\cite{Adhikari:2017esn}. The experiment,
which began data taking in 2016, utilizes eight low-background NaI(Tl) scintillating crystals~\cite{cosinebg}
arranged in a 4$\times$2 array, with a total target mass of 106\,kg.  Each crystal is coupled to two photomultiplier
tubes~(PMTs) to measure the amount of deposited energy in the crystal. The  crystal assemblies are immersed in
2,200\,L of liquid scintillator, which allows for the identification and subsequent reduction of radioactive
backgrounds observed in the crystals~\cite{Park:2017jvs}. The liquid scintillator is
surrounded by copper, lead, and plastic scintillators to reduce the background
contribution from external radiation as well as tag cosmic-ray
muons that transit the apparatus~\cite{Prihtiadi:2017inr}. 

With the initial 59.5 live days exposure of COSINE-100, we reported our first WIMP dark matter search
result~\cite{Adhikari:2018ljm} that excluded the 3$\sigma$ region of allowed WIMP masses and cross sections
that were associated with the DAMA/LIBRA-phase1 signal assuming canonical~(isospin-conserving)
spin-independent~(SI) WIMP interactions in the specific context of the standard WIMP galactic halo model~\cite{Savage:2008er}.
Even though DAMA/LIBRA and COSINE-100 use the same NaI(Tl) target, there are differences.
The DAMA/LIBRA signal is an annual modulation effect while the COSINE-100 result is based on
the time averaged spectral shape~\cite{Kang:2019fvz}. Although the first modulation measurements
from ANAIS-112~\cite{Amare:2019jul} and COSINE-100~\cite{Adhikari:2019off} were recently released,
both experiments still need a few more years of exposure
to reach a modulation sensitivity that is sufficient to probe the DAMA/LIBRA
signal directly~\cite{Adhikari:2017esn,Coarasa:2018qzs}.

It is interesting to compare the DAMA/LIBRA annual modulation signal with the time-averaged
rate considering specific models for the WIMP-nucleon interaction. This is especially the case
for the time-averaged NaI(Tl) results from COSINE-100~\cite{Adhikari:2018ljm}. While the DAMA/LIBRA-phase1 results
used a 2\,keVee~(electron equivalent energy) energy threshold, 
the recent phase2 result has a lower threshold of 1\,keVee~\cite{Bernabei:2018yyw}.
The new low-threshold energy signal has a significantly worse goodness-of-fit for the
canonical SI scattering scenario~\cite{Baum:2018ekm,Kang:2018qvz,Herrero-Garcia:2018mky}, suggesting
that an isospin-violating model in which the WIMP-proton coupling is different from the
WIMP-neutron coupling, or a spin-dependent~(SD) interaction model are better suited for WIMP dark matter
interpretations of the signal. 

To make reliable comparison between the time-averaged rate and the annual modulation amplitude, a local distribution of dark matter particles is necessary.
In this paper, we use standard galactic WIMP halo model~\cite{Lewin:1995rx,freese1987} that has the speed distribution associated with the Maxwell Botzmann,
\begin{eqnarray}
  f(\textbf{v},t) = \frac{1}{N_\mathrm{esc}}~e^{-(v+v_\mathrm{E})^2/2\sigma_v^2},
\label{eq:vdist}
\end{eqnarray}
where $N_\mathrm{esc}$ is a normalization constant, $v_E$ is the Earth velocity relative to the WIMP
dark matter, and $\sigma_v$ is the velocity dispersion. The standard halo parameterization is used
with local dark matter density $\rho_\chi = 0.3~\mathrm{GeV/cm}^3$, $v_E = 232~\mathrm{km/s}$,
$\sqrt{2}\sigma_v = 220~\mathrm{km/s}$ and galactic escape velocity $v_\mathrm{esc} = 544~\mathrm{km/s}$.

Astrophysical parameters related with dark matter local distribution have large uncertainties~\cite{Klypin_2011,Green:2017odb,Wu:2019nhd}. The adoption of various possibilities for the dark matter halo structures
typically extends allowed parameter regions, as studied by DAMA/LIBRA~\cite{Bernabei:2019ajy}. If we consider various halo models that allow different modulation fraction to total rate, DAMA/LIBRA allowed regions will not be fully covered by the COSINE-100 data  as examples shown in Refs.~\cite{Kang:2019fvz,Buckley:2019skk}. This can be improved with larger dataset from COSINE-100 and ANAIS-112 in the future and analysis of data for the model independent annual modulations~\cite{Adhikari:2017esn,Coarasa:2018qzs}.
However, it is still interesting to test the situation based on the widely used standard galactic halo model. 

One noticeable issue with the interpretation of the DAMA/LIBRA obervation in terms of WIMP-nucleon
interactions is the value of the nuclear-recoil quenching factor~(QF).  Quenching factors are the
scintillation light yields for sodium and iodine recoils relative to those for $\gamma$/electron-induced
radiation of the same energy. Most previous studies have used QF values reported by the DAMA/NaI collaboration
in 1996~\cite{BERNABEI1996757} (subsequently referred to as DAMA QF values), that were obtained by measuring
the response of NaI(Tl) crystals to nuclear recoils induced by neutrons from a $^{252}$Cf source. 
The measured responses are compared with the simulated  neutron energy spectrum to obtain QF
values with the assumption that they are independent of the energy of the recoiling nuclide: 
for sodium recoil energies between 6.4 and 97\,keVnr~(nuclear recoil energy), \qnadot=0.30$\pm$0.01;
for iodine recoil energies between 22 and 330\,keVnr, \qidot=0.09$\pm$0.01~\cite{BERNABEI1996757}. 
Recently, results from more refined methods for measuring NaI(Tl) QF values that use monochromatic
neutron beams have been reported~\cite{Collar:2013gu,Xu:2015wha,Stiegler:2017kjw,Joo:2018hom}. In
these measurements, the detection of an elastically scattered neutron at a fixed angle relative to
the incoming neutron beam direction provides an unambiguous knowledge of the energy transferred to the
target nuclide. The QF values from these recent determinations differ significantly from the 1996
DAMA QF results, as shown in Fig.~\ref{fig:QFs}.

In this article, allowed regions in WIMP-nucleon cross-section and WIMP mass parameter space
corresponding to the DAMA/LIBRA signal are presented for some of the different possible dark matter
interactions that are discussed in Ref.~\cite{Baum:2018ekm} using
the recently measured \qna and \qi values. 
The DAMA/LIBRA-phase1~\cite{Bernabei:2013xsa} and phase2~\cite{Bernabei:2018yyw} data are used simultaneously for cases where a good quality-of-fit was obtained. 
The allowed regions from the DAMA/LIBRA data are explicitly compared with the 90\% confidence level~(CL) limits
estimated from the 59.5 day COSINE-100 exposure~\cite{Adhikari:2018ljm}. For comparison, the same data are
interpreted using the DAMA QF values.  For all  of the WIMP-nucleon interactions considered here, we find that
the COSINE-100 data excludes the 3$\sigma$ allowed regions associated with the DAMA/LIBRA data
in the context of the standard WIMP galactic halo model. 

\section{Quenching factor model and implications for the interpretation of the DAMA/LIBRA signal}


The electron-equivalent visible energy $E_{ee}$ produced by recoil nuclei in scintillation detector is typically smaller than its true nuclear recoil energy $E_R$. The ratio of $E_{ee}$ to $E_R$, the nuclear recoil quenching factor (QF), has to be externally evaluated in order to interpret results from dark matter search experiments that use scintillating crystal target/detectors. The DAMA/LIBRA collaboration measured QF values for sodium, \qnadot=0.3$\pm$0.01 averaged over 6.4 to 97\,keVnr, and iodine, \qidot=0.09$\pm$0.01 averaged over 22 to 330\,keVnr~\cite{BERNABEI1996757}. Several measurements in literature between 1994 and 2008, using mono-energetic neutrons produced by neutron generators, obtained consistent results as well~\cite{Spooner:1994ca,Gerbier:1998dm,Jagemann:2006p,Simon:2002cw,Chagani:2008in}. 

\begin{figure}[b]
\centering 
\includegraphics[width=0.95\textwidth]{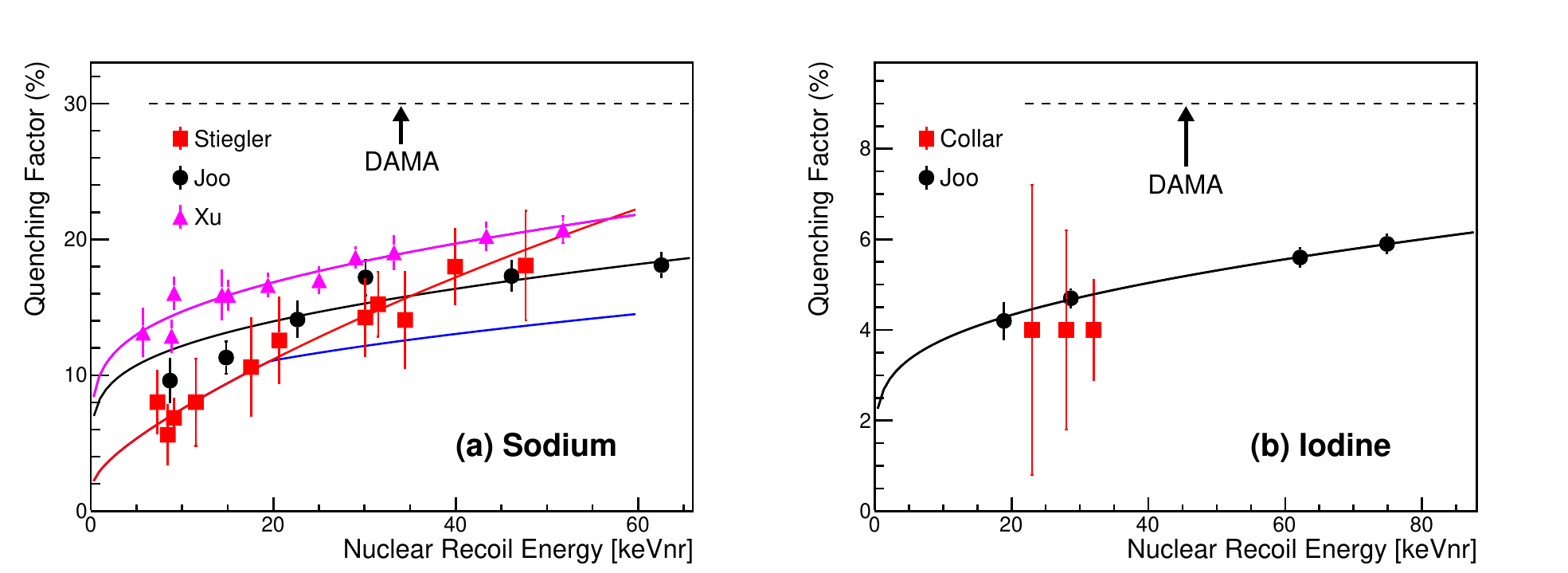}
\caption{
  The nuclear recoil quenching factors of Na~(a) and I~(b) recoils in the NaI(Tl) crystal measured by
  DAMA~\cite{BERNABEI1996757}~(black dashed line) are compared with the recent measurements by
	Stiegler {\it et al.}~\cite{Stiegler:2017kjw}~(red square points), Xu {\it et al.}~\cite{Xu:2015wha}~(magenta triangle points)
  and Joo {\it et al.}~\cite{Joo:2018hom}~(black circle points). 
The new measurements are modeled with an empirical formula based on the
Lindhard~${\it et~al.}$ model~\cite{osti_4701226}. The blue solid line in (a) indicates our assumed 
lower bound of QF systematic uncertainty for $E_R \gtrsim 19.6$\,keVnr that considers the fast
increase of \qna in the Stiegler {\it et al.} data.}
\label{fig:QFs}
\end{figure}

However, recent measurements by Collar~\cite{Collar:2013gu}, Stiegler~${\it et al.}$~\cite{Stiegler:2017kjw}, Xu~${\it et~al.}$~\cite{Xu:2015wha} and Joo~${\it et~al.}$~\cite{Joo:2018hom} reported significantly different results of strong $E_R$ dependence as presented in Fig.\ref{fig:QFs}. 
Main difference of \qna behavior has arised at energy below 20\,keVnr corresponding to approximately 2\,keVee. Efficient noise rejection as well as correct evaluation of trigger and selection efficiencies are essential for proper estimation of the quenching factors in this domain~\cite{Gerbier:1998dm,Collar:2013gu,Xu:2015wha}. 
Considering high light yield crystals and much precise determination of \qna in the new measurements~\cite{Xu:2015wha,Stiegler:2017kjw,Joo:2018hom}, 
we only consider recent four \qna measurements for our modeling. 

In order to parameterize the energy-dependent QF measurements, we use the formula from Lindhard~${\it et~al.}$~\cite{osti_4701226}:
\begin{eqnarray}
  f(E_R) = \frac{kg(\epsilon)}{1+kg(\epsilon)},
\label{eq:qfLindhard}
\end{eqnarray}
where $\epsilon = 11.5Z^{-7/3}~E_R$, $k = 0.133Z^{2/3}A^{1/2}$, $Z$ is the number of protons, and $A$ is the number of nucleons. The function $g(\epsilon)$ is given by~\cite{Lewin:1995rx} to be:
\begin{eqnarray}
  g(\epsilon) = 3\epsilon^{0.15} + 0.7\epsilon^{0.6} + \epsilon.
\end{eqnarray}
The direct application of the Lindhard model to the NaI(Tl) crystals provides a poor match to the recently measured QF values.  We, therefore, consider $k=p_0$ and $\epsilon = p_1 E_R$, where $p_0$ and $p_1$ are fit parameters. This modified Lindhard model well describes the recent measurements of \qna and \qi as shown in Fig.~\ref{fig:QFs} and the fit results are shown in Table~\ref{tab:qffit}. For the \qna measurements, we do not directly use Collar's measurement due to its large uncertainties, which are covered by the other measurements. There are two \qi measurements by Collar and Joo~${\it et~al.}$ as shown in Fig.~\ref{fig:QFs}~(b). In order to estimate the \qi model, we use only the results from Joo~${\it et~al.}$, because the measurement by Joo~${\it et~al.}$ covers that by Collar in terms of energy coverages as well as uncertainties.

Even though the new measurements have consistent energy dependence and lower QF values than those measured by DAMA, there are some mutual differences. These may be due to different environmental conditions such as temperature~\cite{IANAKIEV2009432}, analysis methods
(including different charge integration windows), and different thallium doping concentration of the crystals used for the measurements. In applying these new QF values to the DAMA/LIBRA data, we consider these variations
as a source of systematic uncertainty.  The Joo~{\it et al.}~\cite{Joo:2018hom} results are taken as the central value with allowed systematical variations that span the range between the Xu~{\it et al.}~\cite{Xu:2015wha} and Stiegler~{\it et al.}~\cite{Stiegler:2017kjw} measurements. Figure~\ref{fig:QFs}~(a) shows the three new QF measurement sets, each with its own fit based on the modified Lindhard model. Because of the fast increase of \qna in the Stiegler~{\it et al.} measurements at energies higher than 19.6\,keVnr, the lower bound of systematic uncertainties, denoted by a blue solid line in Fig.~\ref{fig:QFs}~(a), was taken to be the difference between the Xu~{\it et al.} and Joo~{\it et al.} measurements.  In the case of the COSINE-100 data, the Joo~{\it et al.} results were used because these measurements used a crystal from the same ingot, the same data aquisition system~\cite{Adhikari:2018fpo}, and the same analysis framework as the COSINE-100 experimental data.

\begin{table}[t]
\centering
\begin{tabular}{cccc}
\hline
\multicolumn{2}{c}{{\bf Measurement}} & $p_0$ & $p_1$ \\
\hline\hline
\multirow{3}{*}{{\bf Sodium}} & Xu~{\it et al.} & $(7.18\pm1.22)\times10^{-2}$ & $(9.98\pm6.20)\times10^{-3}$ \\
 & Joo~{\it et al.} & $(5.88\pm0.75)\times10^{-2}$ & $(9.12\pm3.16)\times10^{-3}$ \\
 & Stiegler~{\it et al.} & $(9.25\pm5.97)\times10^{-3}$ & $(3.63\pm3.34)\times10^{-1}$ \\
 \hline
{\bf Iodine} & Joo~{\it et al.} & $(1.94\pm0.44)\times10^{-2}$ & $(4.43\pm3.76)\times10^{-3}$ \\
\hline
\end{tabular}
\caption{The fit results of the new QF measurements modeled by modified Lindhard~${\it et~al.}$ model~\cite{osti_4701226} as shown in Fig.~\ref{fig:QFs}.}
\label{tab:qffit}
\end{table}

\begin{figure}[t]
\centering 
\includegraphics[width=0.45\textwidth]{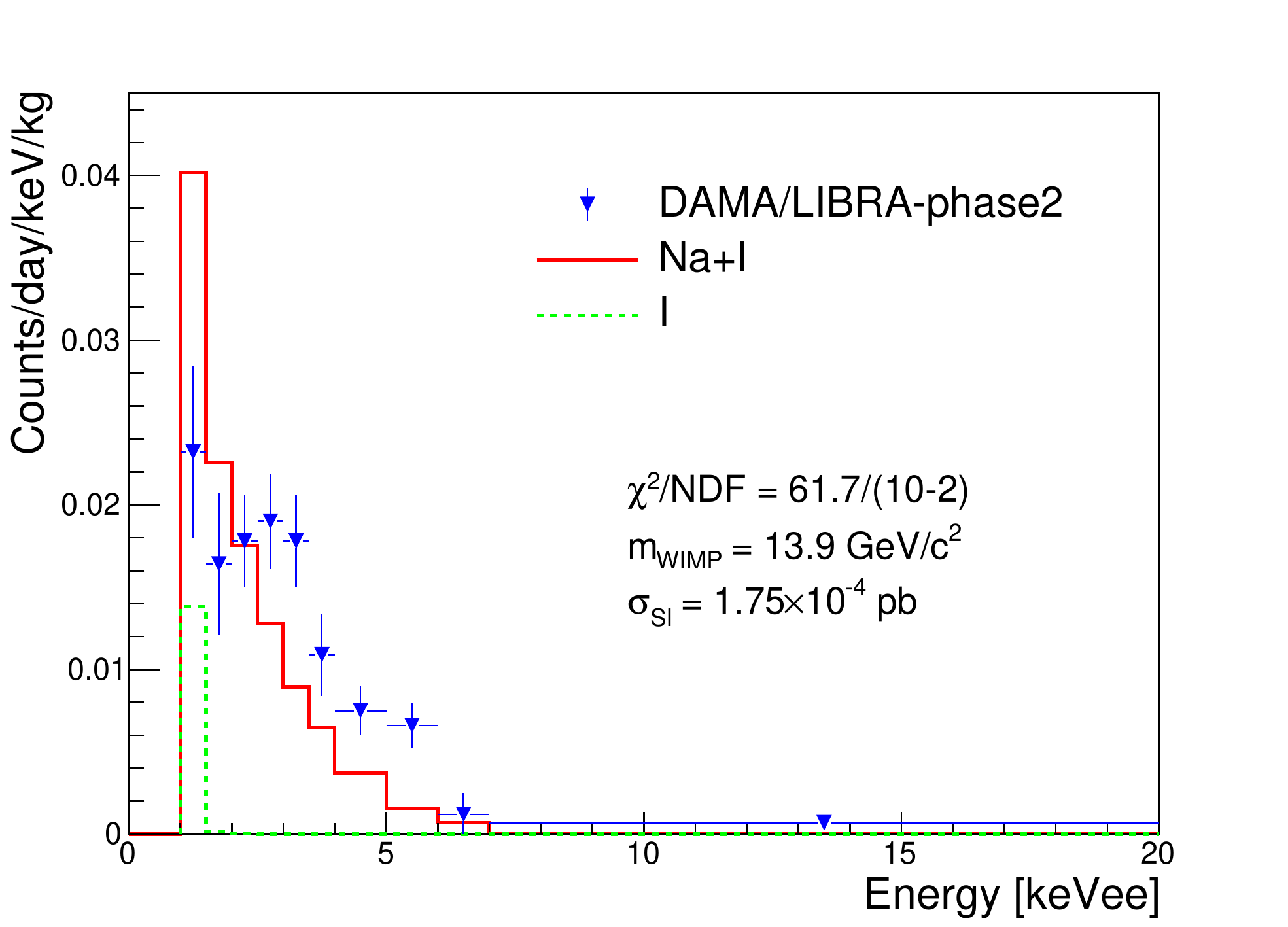}
\includegraphics[width=0.45\textwidth]{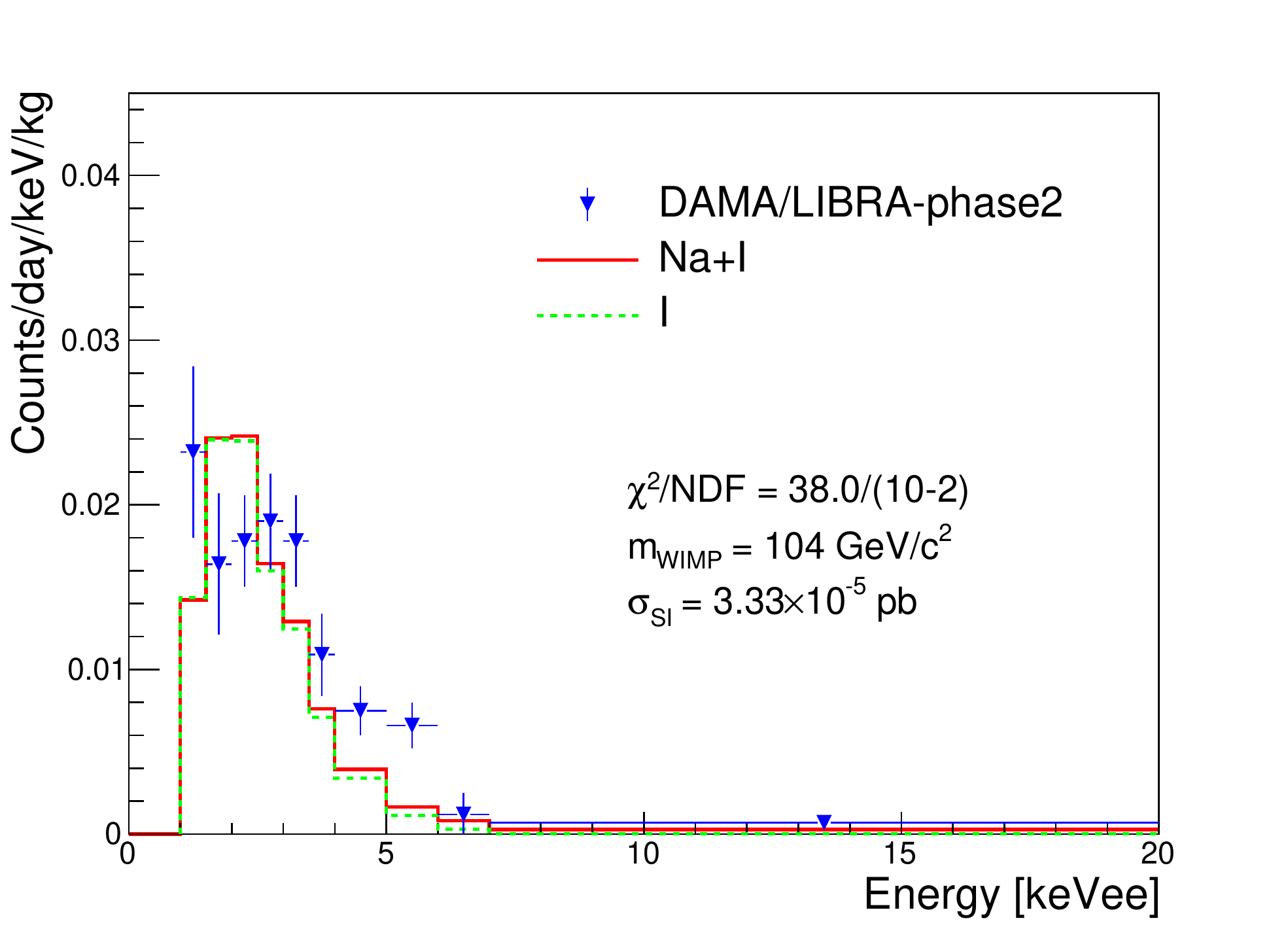}
\includegraphics[width=0.45\textwidth]{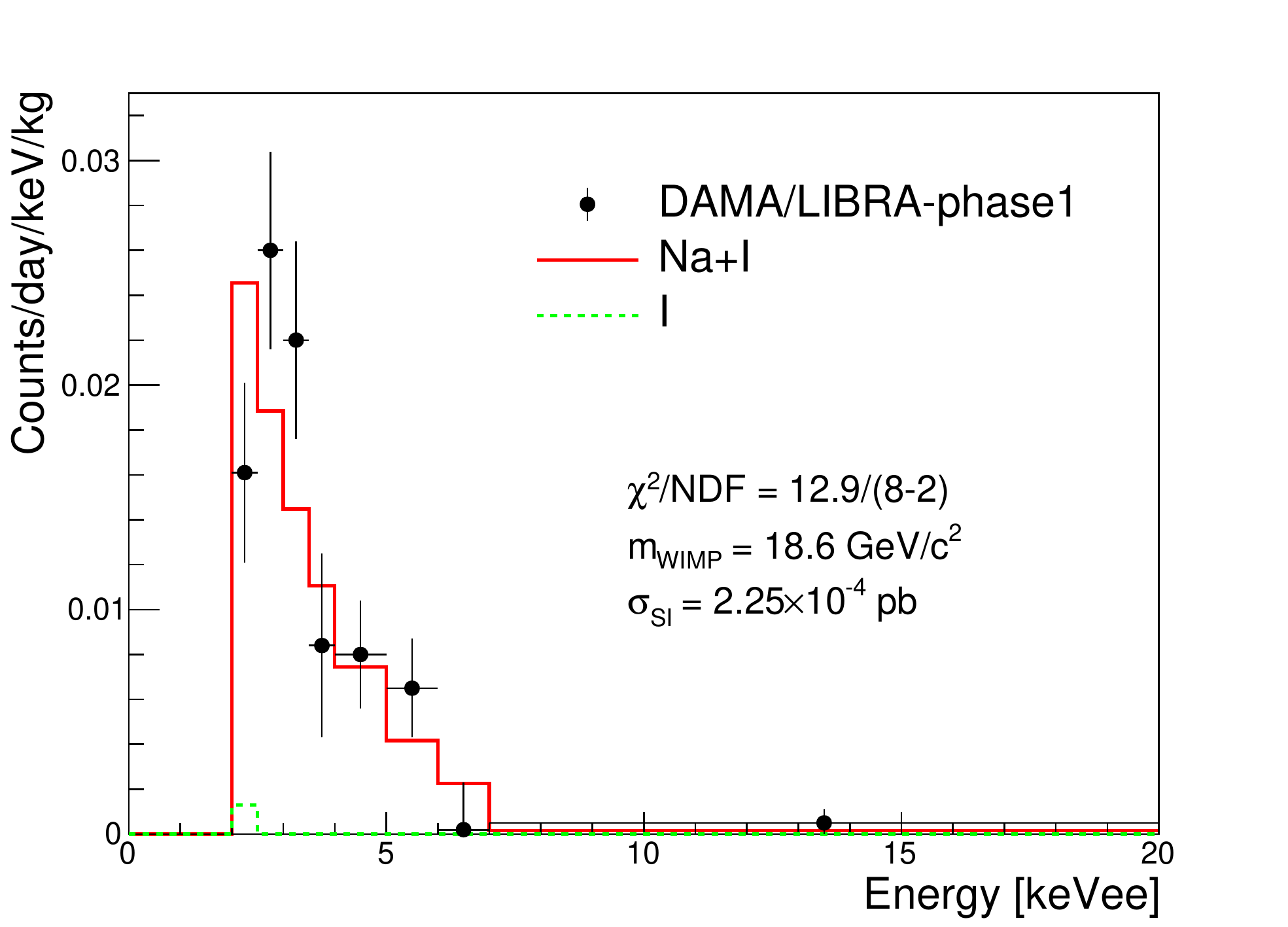}
\includegraphics[width=0.45\textwidth]{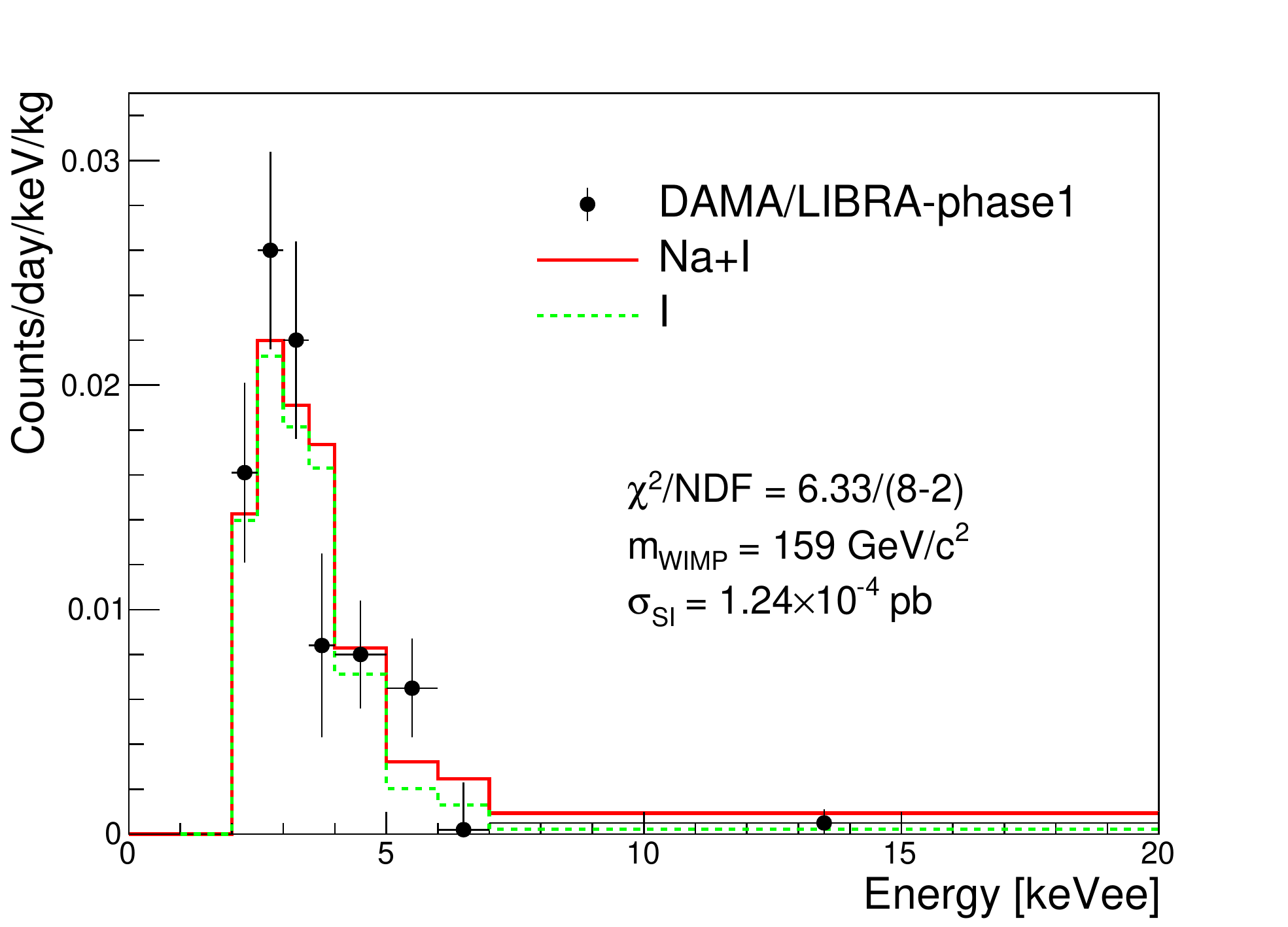}
\caption{
  DAMA/LIBRA's modulation amplitudes~(phase2:top and phase1:bottom) as a function of measured electron-equivalent
  energy are presented for low-mass regions~(left) and high-mass regions~(right) with the best fit models~(red
  solid line), with the
  assumption of canonical SI WIMP interactions and the new QF values. The iodine-only component is denoted by the
  green-dashed line. The DAMA/LIBRA-phase1 data provide good fits for both low-mass and high-mass regions,
  while the phase2 data has large chi-squared values at the best fit points. 
}
\label{fig:SI_spec}
\end{figure}

We use the modulation amplitude results from DAMA/LIBRA-phase1~\cite{Bernabei:2013xsa} and phase2~\cite{Bernabei:2018yyw}, as rebinned in Ref.~\cite{Baum:2018ekm} and shown in Fig.~\ref{fig:SI_spec}. We built a $\chi^2$ fitter to test the DAMA/LIBRA data against the modulation amplitude that is expected for the WIMP interaction under consideration. The energy resolution of the DAMA/LIBRA detector was taken from Refs.~\cite{Bernabei:2008yh,Bernabei:2012zzb}; the reported  DAMA/LIBRA data is efficiency corrected. In order to obtain allowed regions in the WIMP mass vs. WIMP-proton cross-section parameter space, we implement a maximum likelihood method based on the likelihood ratio to fit for mass and cross section values. Confidence regions in these parameters are determined by examining variations of the likelihood values from their maxima. 

\section{Isospin-conserving spin-independent interaction}

\begin{figure}[b]
\centering 
\includegraphics[width=0.8\textwidth]{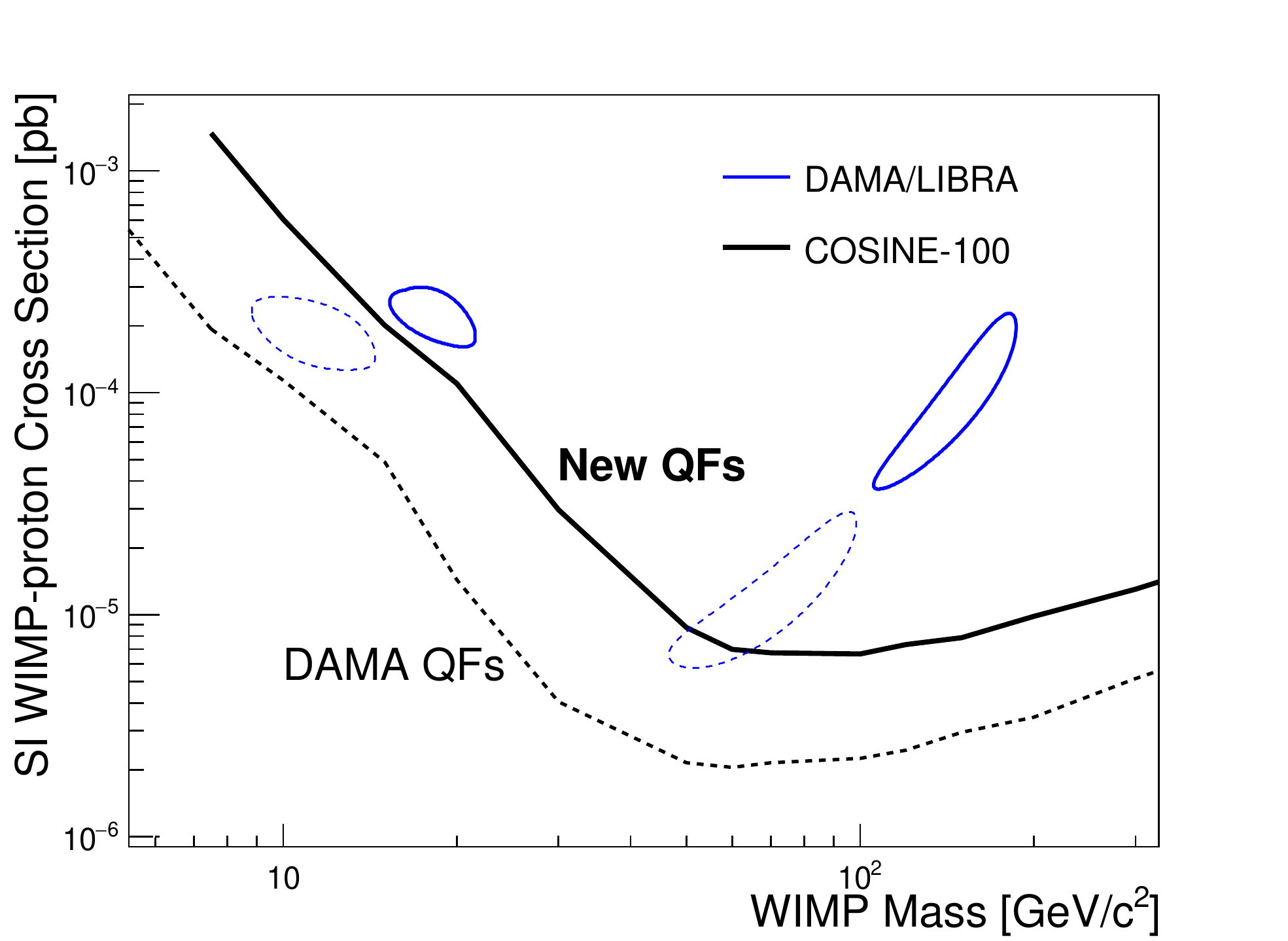}
\caption{ 
  The 3$\sigma$ allowed regions of the WIMP mass and the WIMP-proton cross-section associated with the
  DAMA/LIBRA-phase1 data~(blue solid contours) are compared with the 90\% CL upper limit from the
  COSINE-100 data~(black solid line) with the new QF values. To illustrate the effects of the QF changes,
  we present the 3$\sigma$ regions of the DAMA/LIBRA-phase1 data~(blue dashed contours) and 90\% CL limit
  of the COSINE-100 data~(black dashed line) using the DAMA QF values (from
  Ref.~\cite{Adhikari:2018ljm}).
}
\label{fig:SI_DAMA1}
\end{figure}

For the DAMA/LIBRA-phase1 data, the isospin conserving SI scattering with DAMA QF values provided a
good fit for WIMPs~\cite{Savage:2008er}. On the other hand, the observed DAMA/LIBRA-phase2 modulation data does
not provide a good fit to the expectations for this model ~\cite{Baum:2018ekm,Kang:2018qvz,Herrero-Garcia:2018mky}. 
Switching to the new QF values for both the phase1 and phase2 data does not improve the phase2 data's agreement with
the model, as shown in Fig.~\ref{fig:SI_spec} and summarized in Table~\ref{tab:RESULT}.
As discussed in Ref.~\cite{Baum:2018ekm}, modulation amplitude in the low-WIMP-mass allowed region, which is dominated by WIMP-sodium scattering,
is expected to increase rapidly for recoil energies below 1.5\,keVee because of the onset of contributions from
WIMP-iodine scattering. On the other hand, modulation amplitude in the the high-WIMP-mass allowed region, which is dominated by WIMP-iodine
scattering, is expected to decrease at energies below 1.5\,keVee. Since the DAMA/LIBRA-phase2 data displays a
modulation amplitude that smoothly increases with energy below 1.5\,keVee, the canonical SI WIMP interaction
cannot provide a good fit to the phase2 data.  We, therefore, only use the phase1 data for the interpretation of
the canonical SI WIMP scattering with the new QF values.  As shown in Fig.~\ref{fig:SI_DAMA1}, the best fit
regions of the DAMA/LIBRA-phase1 data with the new QF results show significantly increased values for both the
allowed WIMP masses and WIMP-nucleon cross-sections.  We find that the local minimum value of chi-squared with
the new QF values in the low-mass region increases somewhat,
while the chi-squared value for the high-mass region decreases, as summarized in Table~\ref{tab:RESULT}.

The 90\% confidence level~(CL) upper limits for the COSINE-100 data
are determined using the Bayesian method described in Ref.~\cite{Adhikari:2018ljm}. 
Even though the allowed parameter space from the DAMA/LIBRA-phase1 data is changed by the new QF values, 
the COSINE-100 results still exclude the DAMA 3$\sigma$ region as shown in Fig.~\ref{fig:SI_DAMA1}. 
This is because the dependence on QF values is nearly the same for the DAMA/LIBRA and COSINE-100 measurements. 

\section{Isospin violating spin-independent interaction}

\begin{figure}[t]
\centering 
\includegraphics[width=0.95\textwidth]{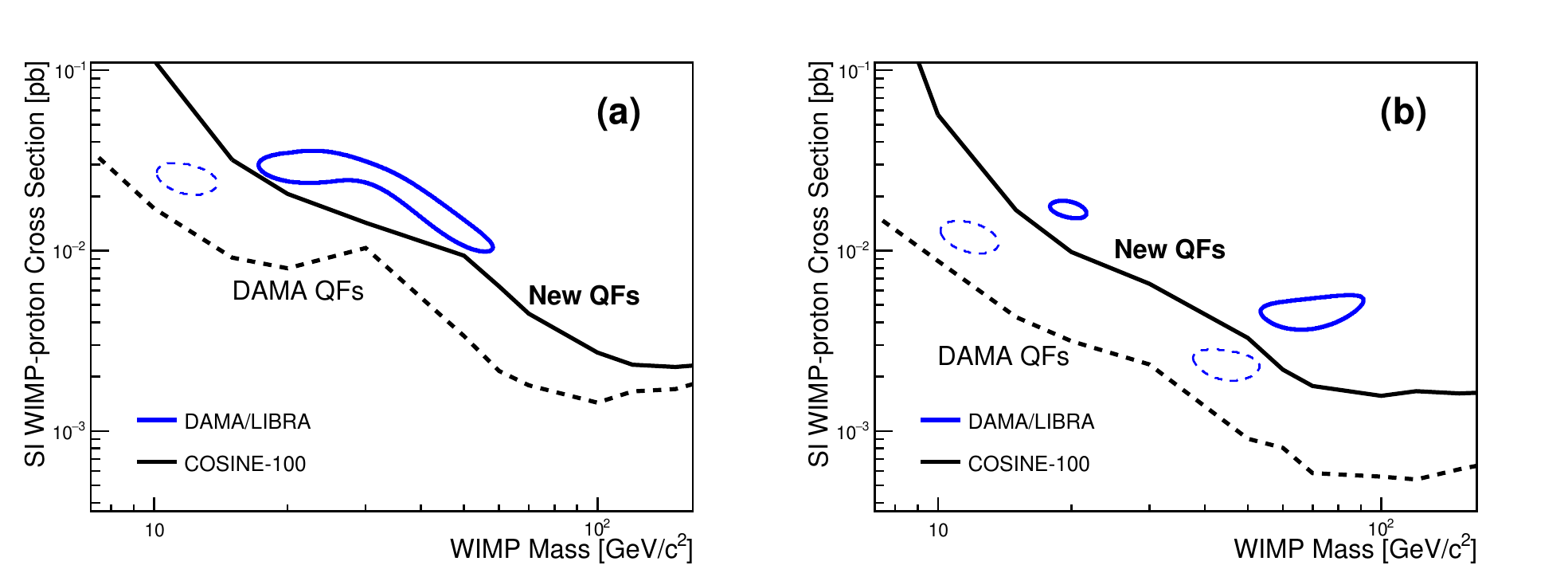}
\caption{ 
  The 3$\sigma$ allowed regions of the WIMP mass and the cross-section associated with the DAMA/LIBRA-phase1+phase2
  data~(blue solid contour) are compared with the 90\% CL upper limit from the COSINE-100 data~(black solid line). 
 The dashed curves shown the results using the DAMA QF values. 
 In each plot, we fix the effective coupling ratios to neutrons and protons $f_n/f_p$ to their best fit values:
 (a) $f_n/f_p=-0.758$ (-0.756) for the low-mass regions and new (DAMA) QF values; 
 (b) $f_n/f_p=-0.712$ (-0.684) for the high-mass regions and new (DAMA) QF values. 
}
\label{fig:ISV_DAMA12}
\end{figure}

It is clear from the above disussion that in order to fit both the DAMA/LIBRA-phase1 and phase2
data~(DAMA/LIBRA-phase1+phase2 data), the contributions from WIMP-iodine scattering have to be
suppressed.  This can be accomplished if the WIMP-proton coupling is different from the WIMP-neutron
coupling~(isospin violating interaction)~\cite{Baum:2018ekm,Kang:2018qvz}.  (Sodium has nearly equal numbers of
protons~(11) and neutrons~(12); iodine has 74~neutrons and 53~protons.)  In this case, three 
parameters are used to fit the DAMA/LIBRA data: the WIMP mass, the WIMP-proton scattering cross-section,
and the ratio between the effective coupling of WIMPs to neutrons and to protons~($f_n/f_p$).
Figure~\ref{fig:ISV_DAMA12} shows the 3$\sigma$-allowed WIMP mass vs. cross-sections regions for the
DAMA/LIBRA-phase1+phase2 data with the new QF values for the best fit values of $f_n/f_p=-0.758$~(a) in the low-mass
and $f_n/f_p=-0.712$~(b) in the high-mass regions. The low-mass and high-mass local minima are significantly shifted
with respect to the results using the DAMA QF values.  The minimum chi-squared values with the new QF values,
listed in Table~\ref{tab:RESULT}, indicate that this model provides a good description of the full
DAMA/LIBRA-phase1+phase2 data set.

The 90\% CL upper limits evaluated from the COSINE-100 data with $f_n/f_p$ values determined from
the best fit to the DAMA/LIBRA-phase1+phase2 data, shown in Fig.~\ref{fig:ISV_DAMA12},
exclude the allowed 3$\sigma$ regions from the DAMA/LIBRA data.
In a scan of different $f_n/f_p$ values over the [-1,1] interval, we find the limits obtained from the COSINE-100
exclude the DAMA/LIBRA allowed 3$\sigma$ regions for all cases.

\section{Spin-dependent interaction}

We use the effective field theory treatment and nuclear form factors from
Ref.~\cite{Fitzpatrick:2012ix,Anand:2013yka,Gresham:2014vja} to estimate the DAMA/LIBRA allowed regions for
spin-dependent~(SD) interactions using the publicly available {\sc dmdd}
package~\cite{dmdd,Gluscevic:2015sqa}. In the fit to the DAMA/LIBRA data, we vary two parameters: the WIMP-mass
and the WIMP-nucleon SD interaction cross section for four cases in terms of ratio between WIMP-neutron and
WIMP-proton SD couplings $a_n/a_p$ (WIMP-proton/neutron only and $a_n/a_p = \pm1$).

\begin{figure}[b]
\centering 
\includegraphics[width=0.95\textwidth]{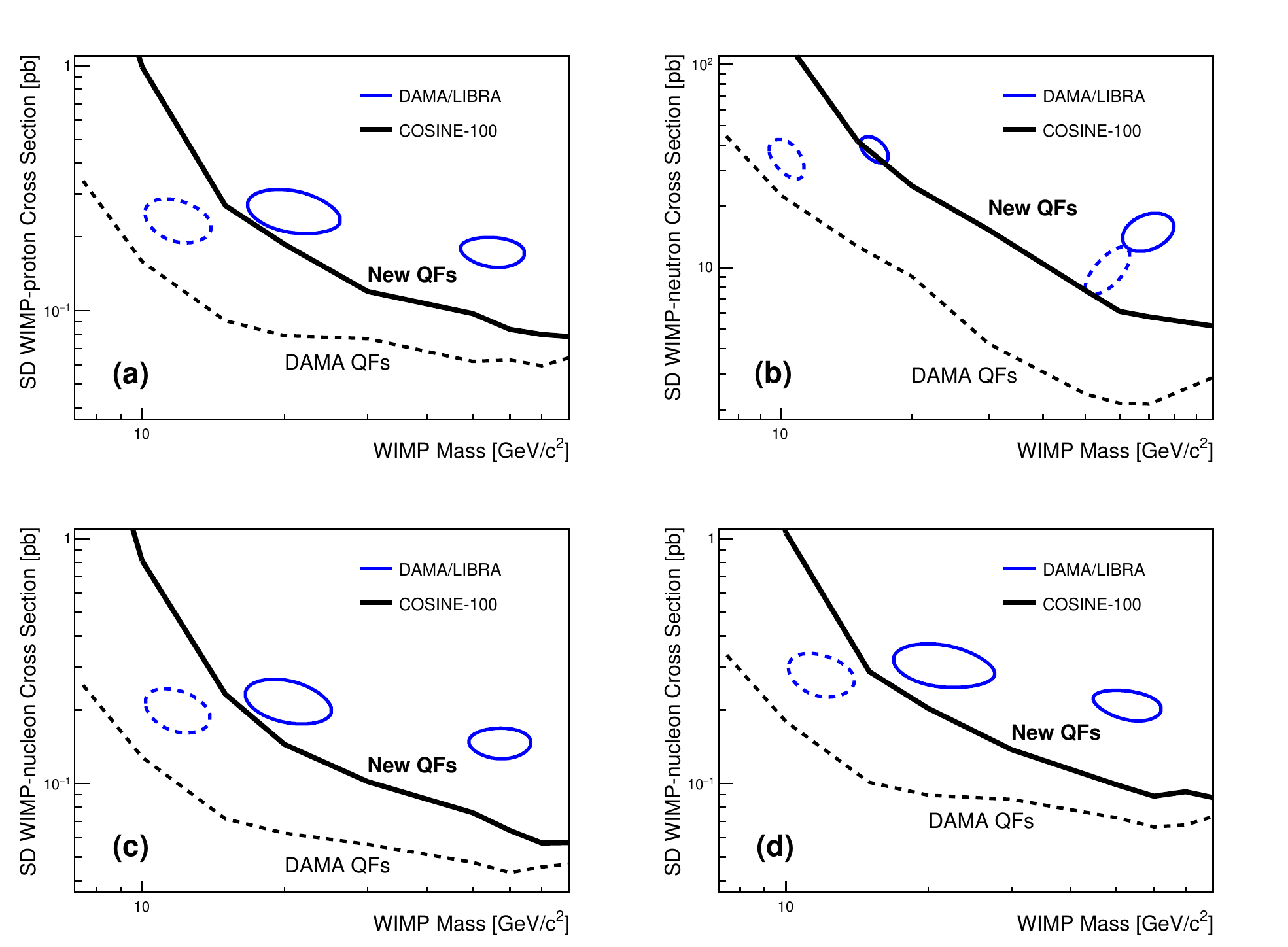}
\caption{ 
  The 3$\sigma$ allowed regions of WIMP mass vs. WIMP-nucleon SD cross-section associated with the
  DAMA/LIBRA-phase1+phase2 data~(blue solid coutours) are compared with the 90\% CL upper limit from the
  COSINE-100 data~(black solid lines). These results use the new QF values; the dashed curves show the results
  using the DAMA QF values. (a) proton only (b) neutron only (c) $a_n/a_p = +1$ (d) $a_n/a_p = -1$.
}
\label{fig:SD_DAMA12}
\end{figure}

In case of the WIMP-neutron only SD interaction~($a_p$=0), the observed DAMA/LIBRA modulation data does not provide a good fit
as shown in Table~\ref{tab:RESULT}. In Fig.~\ref{fig:SD_DAMA12}~(b) it is drawn for the completeness based on likelihood ratio. 
On the other hand, two local minima are obtained with the new QF values for the SD WIMP-proton interaction and other two mixed couplings,
while only a low-mass WIMP has a good fit for the DAMA QF values as shown in Fig.~\ref{fig:SD_DAMA12}. 
However, the chi-squared value of the best fit using the new QF values is slightly worse, as shown in
Table~\ref{tab:RESULT}.  In the high-mass region, the relatively large chi-squared value with the new QF
values corresponds to a similar trend seen in the fit that uses the DAMA QF values. 

Figure~\ref{fig:SD_DAMA12} shows the 90\% CL upper limits obtained from the COSINE-100 data with
the same effective field theory treatment and nuclear form factors for sodium and iodine. The
DAMA/LIBRA 3$\sigma$ allowed regions for SD WIMP-proton interaction hypothesis are excluded by
the 90\%~CL upper limit from the COSINE-100 data. 

\begin{table}[b]
\centering
\begin{tabular}{cccccc}
\hline
\multirow{2}{*}{~~~~~~~Model~~~~~~~} & \multirow{2}{*}{~~~~~~~QF~~~~~~~}
& \multirow{2}{*}{~~~$\chi^2$/NDF~~~} & ~~~$m_\mathrm{WIMP}$~~~
& ~~~~~~~$\sigma_\mathrm{WIMP}$~~~~~~~& \multirow{2}{*}{$f_n/f_p$} \\
&&& [$\mathrm{GeV/c^2}$] & [$\mathrm{pb}$]&\\
\hline\hline
& \multirow{2}{*}{{\bf New}} & 12.9/(8-2) (2.0$\sigma$)& 18.6 & $2.25\times10^{-4}$ & \multirow{4}{*}{1.000} \\
Canonical SI && 6.33/(8-2)  (0.9$\sigma$)& 159 & $1.24\times10^{-4}$&\\
(Phase1 only) & \multirow{2}{*}{DAMA} & 9.62/(8-2)  (1.5$\sigma$)& 11.3 & $1.96\times10^{-4}$ &\\
&& 7.51/(8-2)  (1.1$\sigma$)& 75.5 & $1.40\times10^{-5}$&\\
\hline
& \multirow{2}{*}{{\bf New}} & 61.7/(10-2)  (6.0$\sigma$)& 13.9 & $1.75\times10^{-4}$ & \multirow{4}{*}{1.000} \\
Canonical SI && 38.0/(10-2)  (4.5$\sigma$)& 104 & $3.33\times10^{-5}$&\\
(Phase2 only) & \multirow{2}{*}{DAMA} & 51.6/(10-2) (5.6$\sigma$)& 8.96 & $1.61\times10^{-4}$ &\\
&& 20.2/(10-2) (2.6$\sigma$)& 59.6 & $8.41\times10^{-6}$&\\
\hline
& \multirow{2}{*}{{\bf New}} & 19.4/(18-3)  (1.3$\sigma$)& 19.5 & $2.90\times10^{-2}$&-0.758\\
Isospin violating SI&& 17.5/(18-3)  (1.1$\sigma$)& 69.2 & $4.55\times10^{-3}$&-0.712\\
(Phase1+2) & \multirow{2}{*}{DAMA} & 17.1/(18-3)  (1.0$\sigma$)& 11.8 & $2.54\times10^{-2}$&-0.756\\
  && 17.4/(18-3)  (1.0$\sigma$)& 44.6 & $2.36\times10^{-3}$&-0.684\\
\hline
& \multirow{2}{*}{{\bf New}} & 24.2/(18-2)  (1.7$\sigma$)& 20.7 & $2.59\times10^{-1}$& \multirow{4}{*}{-}\\
WIMP-proton SD && 30.5/(18-2)  (2.4$\sigma$)& 55.6 & $1.74\times10^{-1}$&\\
(Phase1+2) & \multirow{2}{*}{DAMA} & 17.3/(18-2)  (0.9$\sigma$)& 11.8 & $2.37\times10^{-1}$&\\
&& 45.2/(18-2)  (3.8$\sigma$)& 42.3 & $1.55\times10^{-1}$&\\
\hline
& \multirow{2}{*}{{\bf New}} & 50.8/(18-2)  (4.3$\sigma$)& 16.4 & $3.83\times10$& \multirow{4}{*}{-}\\
WIMP-neutron SD && 44.0/(18-2)  (3.7$\sigma$)& 69.6 & $1.52\times10$&\\
(Phase1+2) & \multirow{2}{*}{DAMA} & 37.5/(18-2)  (3.1$\sigma$)& 10.4 & $3.50\times10$&\\
&& 36.6/(18-2)  (3.0$\sigma$)& 56.7 & $9.89$&\\
\hline
& \multirow{2}{*}{{\bf New}} & 25.8/(18-2)  (1.9$\sigma$)& 20.2 & $2.20\times10^{-1}$& \multirow{4}{*}{-}\\
mixed SD: $a_n = a_p$ && 31.7/(18-2)  (2.5$\sigma$)& 57.4 & $1.47\times10^{-1}$&\\
(Phase1+2)& \multirow{2}{*}{DAMA} & 17.2/(18-2)  (0.9$\sigma$)& 11.8 & $2.02\times10^{-1}$&\\
&& 41.2/(18-2)  (3.5$\sigma$)& 44.1 & $1.24\times10^{-1}$&\\
\hline
& \multirow{2}{*}{{\bf New}} & 22.8/(18-2)  (1.6$\sigma$)& 21.2 & $3.09\times10^{-1}$& \multirow{4}{*}{-}\\
mixed SD: $a_n = -a_p$&& 29.2/(18-2)  (2.3$\sigma$)& 53.4 & $2.10\times10^{-1}$&\\
(Phase1+2)& \multirow{2}{*}{DAMA} & 17.6/(18-2)  (0.9$\sigma$)& 11.8 & $2.81\times10^{-1}$&\\
&& 52.2/(18-2)  (3.4$\sigma$)& 40.4 & $1.98\times10^{-1}$&\\
\hline
\end{tabular}
\caption{ 
  The best fit values for the comparison of six WIMP-nucleon interaction hypotheses to the DAMA/LIBRA data
  are summarized.
  Here we present the fit results based on both the DAMA and new QF values. 
  The first and second groups of rows are for the canonical SI interaction using the phase1 and phase2 data,
  respectively. The other groups use the DAMA/LIBRA-phase1+phase2 data for the fit. 
	The third group is for the isospin-violating SI interaction while the next four groups
  are for the SD interactions. 
	The SD interactions are shown for proton only interaction~(fifth group), neutron only interaction~(sixth group), and mixed couplings of $a_n/a_p$=1~(seventh group)
	and $a_n/a_p=-1$~(eight group). 
  The canonical SI interaction for the DAMA/LIBRA-phase2 data and neutron only SD interaction do not provide good fits, while for the other
  cases good fits are obtained. 
}
\label{tab:RESULT}
\end{table}

\section{Discussion}
We examine the compatibility of the DAMA/LIBRA and COSINE 100 data in the context of various WIMP
dark matter interaction hypotheses and taking into account the recently measured nuclear recoil QF values for sodium and iodine. 
Here we assume the standard galatic WIMP halo model with astrophysical parameters: $\rho_\chi = 0.3~\mathrm{GeV/cm}^3$, $v_E = 232~\mathrm{km/s}$,
$\sqrt{2}\sigma_v = 220~\mathrm{km/s}$, and $v_\mathrm{esc} = 544~\mathrm{km/s}$.
We find that the DAMA/LIBRA-phase2 data are not compatible with canonical SI WIMP interaction in the context of
the standard WIMP galactic halo model using the new QF values.  Moreover, the DAMA/LIBRA-phase1 data only are well
fitted but with significant shifts in both the allowed WIMP-mass and WIMP-nucleon cross-section values.
We successfully obtained allowed regions from the DAMA/LIBRA-phase1+phase2 data for an isospin-violating
interaction hypothesis, as well as for spin-dependent WIMP-proton and mixed couplings of proton and neutron interactions with the new QF values. However,
for all the WIMP-dark matter interpretations of the DAMA/LIBRA data considered here, the COSINE-100
limits based on the initial 59.5~days' exposure exclude the 3$\sigma$ allowed regions for the DAMA/LIBRA
modulation signal at the 90\% CL. Because the COSINE-100 experiment uses the same NaI(Tl)
target medium as the DAMA/LIBRA experiment, this result strongly constrains models that purport to explain
the DAMA/LIBRA modulation signal as being due to interactions of WIMPs in the galactic dark matter halo with
nuclides in NaI(Tl) crystal detectors. 



\acknowledgments
We thank the Korea Hydro and Nuclear Power (KHNP) Company for providing underground laboratory space at Yangyang.
This work is supported by:  the Institute for Basic Science (IBS) under project code IBS-R016-A1 and NRF-2016R1A2B3008343, Republic of Korea;
UIUC campus research board, the Alfred P. Sloan Foundation Fellowship,
NSF Grants No. PHY-1151795, PHY-1457995, DGE-1122492,
WIPAC, the Wisconsin Alumni Research Foundation, United States; 
STFC Grant ST/N000277/1 and ST/K001337/1, United Kingdom;
and Grant No. 2017/02952-0 FAPESP, CAPES Finance Code 001, Brazil.

\bibliographystyle{JHEP}

\providecommand{\href}[2]{#2}\begingroup\raggedright\endgroup

\end{document}